\documentclass[
reprint,
superscriptaddress,
amsmath,amssymb,
prl
]{revtex4-2}

\usepackage{graphicx}% Include figure files
\usepackage{dcolumn}% Align table columns on decimal point
\usepackage{bm}% bold math
\usepackage{hyperref}% add hypertext capabilities
\usepackage[dvipsnames]{xcolor}
\usepackage{soul}
\usepackage{ulem}
\usepackage[left]{lineno}
\usepackage{braket}

\begin{document}
%\preprint{APS/123-QED}
\title{Line lasing in a two-dimensional lattice of orbital photonic resonators}

\author{Tony~Mathew~Blessan}
\affiliation{Univ. Lille, CNRS, UMR 8523 -- PhLAM -- Physique des Lasers Atomes et Mol\'ecules, F-59000 Lille, France}

\author{Bastian~Real}
\affiliation{Univ. Lille, CNRS, UMR 8523 -- PhLAM -- Physique des Lasers Atomes et Mol\'ecules, F-59000 Lille, France}

\author{Marijana~Milicevic}
\affiliation{Université Paris-Saclay, CNRS, Centre de Nanosciences et de Nanotechnologies, 91120 Palaiseau,France}

\author{Isabelle~Sagnes}
\affiliation{Université Paris-Saclay, CNRS, Centre de Nanosciences et de Nanotechnologies, 91120 Palaiseau,France}

\author{Aristide~Lemaître}
\affiliation{Université Paris-Saclay, CNRS, Centre de Nanosciences et de Nanotechnologies, 91120 Palaiseau,France}

\author{Luc~Le~Gratiet}
\affiliation{Université Paris-Saclay, CNRS, Centre de Nanosciences et de Nanotechnologies, 91120 Palaiseau,France}

\author{Abdelmounaim~Harouri}
\affiliation{Université Paris-Saclay, CNRS, Centre de Nanosciences et de Nanotechnologies, 91120 Palaiseau,France}

\author{Sylvain~Ravets}
\affiliation{Université Paris-Saclay, CNRS, Centre de Nanosciences et de Nanotechnologies, 91120 Palaiseau,France}

\author{Jacqueline~Bloch}
\affiliation{Université Paris-Saclay, CNRS, Centre de Nanosciences et de Nanotechnologies, 91120 Palaiseau,France}

\author{Clément~Hainaut}
\affiliation{Univ. Lille, CNRS, UMR 8523 -- PhLAM -- Physique des Lasers Atomes et Mol\'ecules, F-59000 Lille, France}

\author{Alberto~Amo}
\email{alberto.amo-garcia@univ-lille.fr} 
\affiliation{Univ. Lille, CNRS, UMR 8523 -- PhLAM -- Physique des Lasers Atomes et Mol\'ecules, F-59000 Lille, France}

%\date{\today}

\begin{abstract} 
The engineering of specialty lasers with unconventional mode structures is one of the modern challenges in the development of integrated coherent sources. Examples include the use of bound states in the continuum, microlasers with orbital angular momentum, Dirac-band lasers and topological lasers. In this work we engineer a two-dimensional lattice of coupled micropillars with lasing line modes. We use a convenient combination of orbital photonic modes to design photonic bands which are flat in one direction and dispersive in the perpendicular one giving rise to line lasing modes. Such an architecture opens the possibility of implementing densely packed lasing matrices in compact two dimensional lattices. 

\end{abstract}
\maketitle

%%%%%%%%%%%%%%%%%%%%%%%%%%  body  %%%%%%%%%%%%%%%%%%%%%%%%%%
\section{Introduction}
The use of periodic photonic structures has been one of the main pathways to design specialty semiconductor lasers with small sizes and straight integrability. 
Early photonic crystal lasers introduced defect modes to produce single mode out-of-plane laser emission in extremely compact designs~\cite{painter_two-dimensional_1999}. 
Suffering from limited output power, there have been important efforts to implement designs with large area, single mode operation in integrated geometries with high directionality. 

A successful pathway has been the use of double-lattice photonic crystal resonators based on an intricate interplay of interference and Hermitian and non-Hermitian couplings~\cite{inoue_general_2022}. 
A realization based on AlGaAs materials has demonstrated several Watts output power with submilimeter footprints~\cite{yoshida_double-lattice_2019}. 
A different approach is to engineer photonic band structures based on Dirac dispersions~\cite{bravo-abad_enabling_2012}. 
This method exploits the inherently large frequency separation between in-plane modes in linear dispersions to achieve large-area single-mode operation~\cite{contractor_scalable_2022, yang_topological-cavity_2022, ma_room-temperature_2023}.
Another strategy is to use the one-dimensional edge state of a two-dimensional lattice whose bands possess a non-trivial topology. 
The robustness of the topological edge mode to local disorder ensures the mode locking of a large number of resonators, resulting in powerful output emission with interesting directional and orbital angular momentum properties both for in-plane~\cite{Bahari2017,Bandres2018,Zeng2020}, out-of-plane operation~\cite{Bahari2021,Dikopoltsev2021} and lasing in line interface modes~\cite{bennenhei_organic_2024}.

In this article, we propose and demonstrate an architecture that enables a novel type of lasing geometry based on a lattice of coupled AlGaAs micropillars displaying polariton emission. 
Specifically, we report lasing from line modes in a two-dimensional Lieb lattice that combines micropillars with different orbital photonic modes.
The coupling of orbital modes of individual resonators has been extensively used in photonic~\cite{Jacqmin2014,Cantillano2017, zhang_realization_2023, vicencio_multi-orbital_2025} and atomic systems~\cite{Wirth2011, li_topological_2013, beugeling_topological_2015} to engineer elaborate band structures.
Our design is based on coupled micropillars that combine \textit{s} and \textit{p} type orbitals to create photonic bands with a flat dispersion along one spatial direction and a dispersive band along the perpendicular direction.
The eigenmodes of this arrangement are independent line modes that cover the whole lattice. 
We observe lasing in such line modes with a reconfigurable location in the lattice determined by the gain profile under optical pumping.
Interestingly, when two line lasers cross, phase locking of the lasers takes place under the proper excitation conditions. 
Simulations based on a model of a pumped reservoir feeding the polariton lattice modes show that coupling between orbital modes induced by an elliptical asymmetry of the micropillars is responsible for the phase locking. 
The architecture demonstrated in this work opens new perspectives on the use of orbital lattices for the implementation of densely packed lasing matrices. 

\begin{figure*}[t!]
\centering\includegraphics[width=1\textwidth]{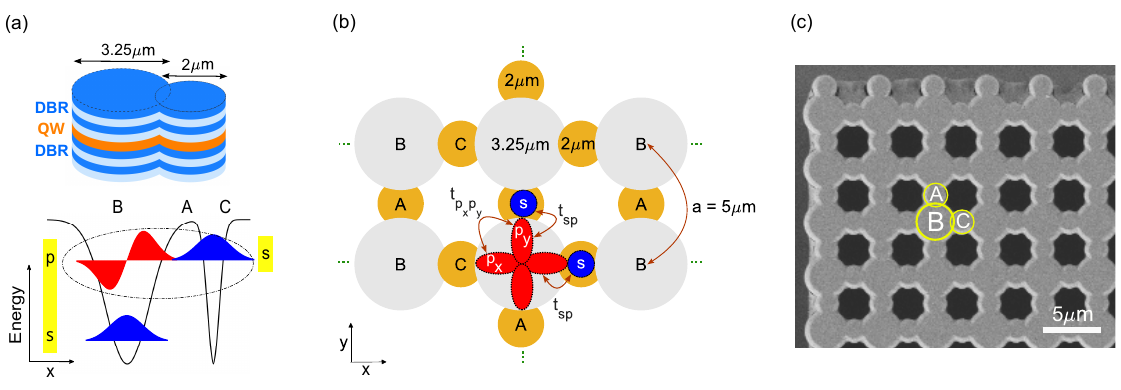}
\caption{(a) Scheme of two coupled micropillars of different diameter. As schematically depicted on the energy level diagram below, the diameters of the micropillars are designed for the $p_x$ and $p_y$ modes of the large micropillar to be at the energy of the $s$ modes of the smaller micropillar. (b) Arrangement of the overlapping orbitals in the $sp$ bands of the Lieb lattice. (c) Scanning electron microscope image of a typical $sp$ Lieb lattice. \label{fig1}}
\end{figure*}

\section{Orbital Lieb lattice}

The building blocks used to implement an orbital Lieb lattice are AlGaAs micropillars grown via molecular beam epitaxy on a two-dimensional GaAs substrate. 
The lower and upper Bragg mirrors are made of 31 and 27 pairs, respectively, of $\lambda/4$ alternating layers of Al$_{0.1}$Ga$_{0.9}$As and Al$_{0.95}$Ga$_{0.05}$As. 
The central GaAs spacer has an optical width of $\lambda$. 
We design $\lambda$ to be close to 849~nm, the lowest energy transition at 4~K of an In$_{0.05}$Ga$_{0.95}$As quantum well of 17~nm in width located at the center of the spacer.

The microcavity wafer is then etched down to the substrate in the form of a Lieb lattice of coupled micropillars (see Fig.~\ref{fig1}(c)).
The three-dimensional confinement of light in each individual micropillar results in a series of discrete photonic modes with $s$, $p$, $d$, ... symmetries (see Figs.~\ref{fig1}(a)-(b)), a labeling inspired by the electronic orbitals of the hydrogen atom~\cite{Galbiati2012,Schneider2017}.
The photon frequency of the modes is determined by the diameter of the micropillar.
The Lieb lattice combines micropillars of two different diameters (3.25 and 2$\mu m$) with a center-to-center interpillar distance of 2.5$\mu$m and a lattice constant of 5$\mu$m (see Fig.~\ref{fig1}).
The size of the pillars is designed such that the $s$ modes of the smallest diameter micropillars are at the energy of the first excited $p_x$ and $p_y$ modes of the large diameter micropillars (Fig.~\ref{fig1}(a)).
%This configuration enables the implementation of a Lieb lattice of overlapping micropillars.

\begin{figure*}[t!]
\centering\includegraphics[width=1\textwidth]{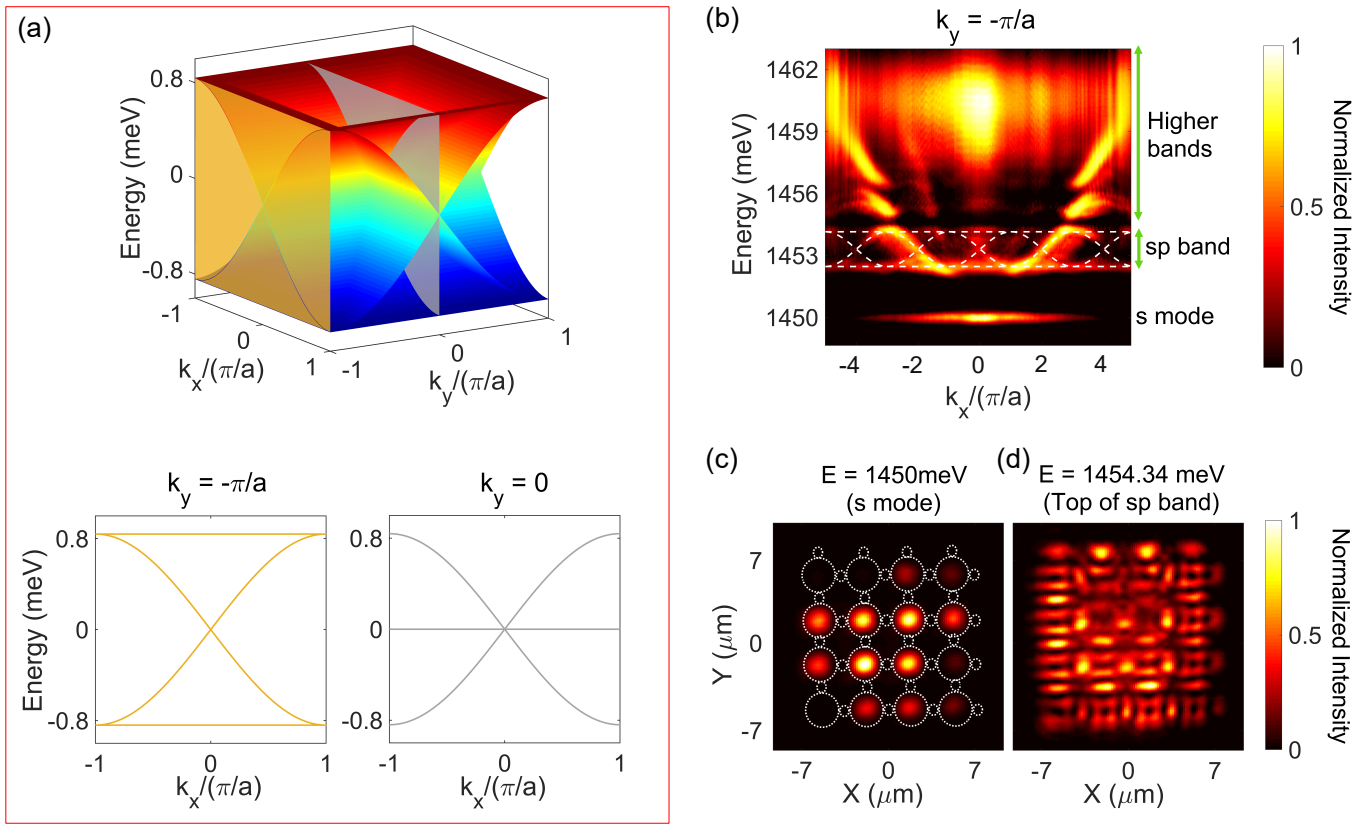}
\caption{(a) Eigenvalues of the $sp$ orbital bands computed from a tight binding model with nearest-neighbors. (b) Measured bands at $k_y = -\pi/a$ at low excitation power. Dashed lines display a fit to the eigenvalues of Eq.~(\ref{eq:H_sp}). (c) Real space emission measured at the energy of the $s$ mode for a $4 \times 4$ lattice. (d) Real space emission measured at the energy of the top of the $sp$ band.\label{fig2}}
\end{figure*}

The lowest energy modes of the structure are the independent $s$ modes of the large micropillars.
The first set of excited bands hybridizes the cylindrical symmetric $s$ orbitals of the small micropillars at A and C sites with the $p_x$ and $p_y$ antisymmetric orbitals of the larger pillars at B sites (Fig.~\ref{fig1}(a) and (b)). 
The eigenenergies and eigenvectors of this band can be described using a tight-binding Hamiltonian:

\begin{widetext}
\begin{equation}
H_{\text{sp}} = \sum_{l,m} -t_{\text{sp}} \left(|\psi_{l,m}^{\text{B},p_y}\rangle\langle\psi_{l,m}^{\text{A},s}|+ |\psi_{l,m}^{\text{B},p_x}\rangle\langle\psi_{l,m}^{\text{C},s}|+|\psi_{l,m}^{\text{A},s}\rangle\langle\psi_{l,m+1}^{\text{B},p_y}|+|\psi_{l,m}^{\text{C},s}\rangle\langle\psi_{l+1,m}^{\text{B},p_x}|\right) +\text{h.c.}
\label{eq:H_sp}
\end{equation}
\end{widetext}
We have used the $s$ orbital energy in the small pillars as the reference energy, and the basis $\{|\psi_{l,m}^{\text{J},\mu}\rangle\}$ of lattice sites that host a single orbital \( \mu = s \) for the J~=~A and C pillars and a pair of orbitals \( \mu \in \{p_x, p_y\} \) for the J~=~B pillars, with $l,m$ labeling the unit-cell positions and $t_{sp}$ the nearest-neighbor hopping amplitude.
The diagonalization of the Hamiltonian in momentum space displays four different bands that cross at the center of the Brillouin zone, see Fig.~\ref{fig2}(a).
Each band is flat along one quasimomentum direction and dispersive along the perpendicular one.

The origin of this intriguing band structure is the directional coupling of the $p_x$ and $p_y$ modes of the large micropillars, displayed in Fig.~\ref{fig1}(b).
At B sites, the $p_y$ orbitals only couple to the $s$ modes of the A sites in the vertical direction, while the $p_x$ orbitals couple only to the C sites in the horizontal direction.
Therefore, the eigenmodes of this lattice are independent one-dimensional line modes along the vertical and horizontal lines of the lattice. 
The one-dimensional nature of the modes results in a band dispersion that is flat along one quasimomentum direction (corresponding to the direction in real space perpendicular to the considered line). 
Along the other quasimomentum direction, the bands follow the typical cosine-like dispersion of a one-dimensional lattice. % with a degeneracy corresponding to the number of lines and rows.

The mechanism for the appearance of the line modes is the directional coupling through the $p_x$ and $p_y$ modes of the B sites.
This is very different to the interference mechanism at the origin of compact localized states in a standard Lieb lattice~\cite{Klembt2017, Whittaker2018,harder_exciton-polaritons_2020, scafirimuto_tunable_2021,alyatkin_quantum_2021, lovett_flat-band_2025}.
The compact localized states involve a minimum of four sites in a square geometry and give rise to a two-dimensional flat band. 
Earlier works in Lieb lattices of micropillars have shown that, both in the $s$ bands and $p$ bands, lasing is dominated by compact localized states highly localized over a few sites only, in a square pattern and emitting at the mid energy of these bands~\cite{Klembt2017,Whittaker2018}.
The crucial novelty in our work is that by mixing $s$ and $p$ orbitals, all the eigenmodes are line modes, which enable the observation of line lasing.

Figure~\ref{fig2}(b) displays the measured low-power photoluminescence of a $sp$ Lieb lattice in momentum space as a function of $k_x$ for $k_y=-\pi/a$ .
A large Gaussian spot at a wavelength of 820nm is used for excitation.
Emission from several bands is evidenced.
The lowest energy one, fully flat, corresponds to the emission from the $s$ modes localized at each of the large B micropillars. 
This is confirmed in Fig.~\ref{fig2}(c), which displays the real space photoluminescence at the energy of the $s$ modes in a Lieb lattice with $4 \times 4$ unit cells.
The two-dimensional real-space images are obtained using standard spectral tomography which spatially and spectrally resolves the photoluminescence of the lattice~\cite{Nardin2009c, Klembt2017, Whittaker2018}.
Well separated above the $s$ modes, the $sp$ set of bands is visible with flat bands at the bottom and top of the set for this specific measurement at $k_y=-\pi/a$.
A fit of the eigenvalues of Eq.~\ref{eq:H_sp} to the measured $sp$ bands in Fig.~\ref{fig2}(b) -- see dashed line --  results in a value of $t_{sp}$ = 0.42meV. 
%Note that the crossing points of the dispersive bands is not located at the middle of energy of the band ensemble as predicted in the tight binding calculations shown in Fig.~\ref{fig2}(a).
%The reason is the presence of next nearest-neighbour coupling between the $s$ modes of A sites of adjacent unit cells and between the $s$ modes of B sites of adjacent unit cells, which does not alter the line geometry of the eigenmodes (see supplementary).
The real-space emission at the energy of the top of the band is shown in Fig.~\ref{fig2}(d), revealing its mixed $s$ and $p$ orbital structure.
At even higher energy, other orbital bands are observed in Fig.~\ref{fig2}(b).

\section{Lasing in line modes}
%\section{Characterization of the orbital Lieb lattice}

\begin{figure*}[t!]  
\centering
\includegraphics[width=\textwidth]{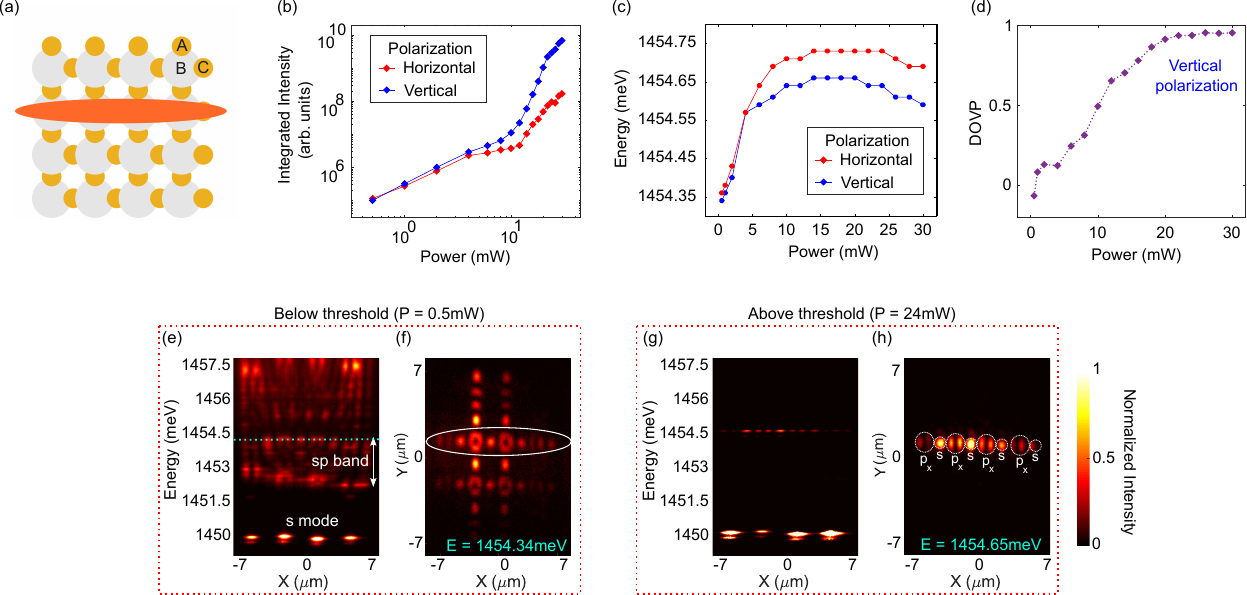}
\caption{(a) Sketch of the configuration for the excitation of an horizontal line of the lattice. (b) Integrated intensity of the upper $sp$ mode along the line in which the lattice is excited. The intensity is resolved in linear polarization. (c) Measured emitted energy of the vertically and horizontally polarized top modes of the $sp$ band. (d) Degree of vertical polarization computed from (b). (e) Emitted spectrum along a line crossing the center of the excitation spot at low excitation power (0.5~mW). (f) Real-space photoluminescence at the energy of the top of the $sp$ band. (g)-(h) Same as (e)-(f) at an excitation power above the lasing threshold. \label{fig3}}
\end{figure*}

We study lasing along different lines of the Lieb lattice by exciting it with an elongated Gaussian spot. The spot, generated with a combination of a cylindrical and an aspherical lens (8~mm focal length for the latter), has full widths at half maxima of 2.5$\mu$m vertically and 18$\mu$m horizontally.
The spot is located at the center of one of the bulk lines of the lattice, sketched in Fig.~\ref{fig3}(a).
The Supplemental Material includes a detailed description of the experimental set-up. 
Energy-resolved photoluminescence is collected in transmission geometry. 

%%%%%%%%%%%%%%%%%%%%%%%%%%%%%%%%%%%%%%%%%%%%%%%%
%%%%%%%%%%%%%%%%%%%%%%%%%%%%%%%%%%%%%%%%%%%%%%%%
%%%%%%%%%%%%%%%%%%%%%%%%%%%%%%%%%%%%%%%%%%%%%%%%
%%%%%%%%%%%%%%%%%%%%%%%%%%%%%%%%%%%%%%%%%%%%%%%%

% \begin{figure}[t!]  
% \centering
% \includegraphics[width=0.7\textwidth]{Fig4.pdf}
% \caption{(a) Integrated intensity of the upper $sp$ mode along the line in which the lattice is excited. The intensity is resolved in linear polarization. (b) Degree of linear polarization computed from (a). \label{fig4}}
% \end{figure}

At low excitation power, Fig.~\ref{fig3}(e), we observe emission from the $s$ modes at 1450~meV, the $sp$ bands in the range 1452.22-1454.34~meV and other orbital bands at higher energy.
Figure~\ref{fig3}(e) shows that at the energy of the top of the $sp$ bands, the emission originates from various vertical and horizontal lines of the lattice.
When the excitation power is increased, the emitted intensity at the energy of the top mode of the $sp$ band displays a lasing threshold of about 10~mW of input power (see Fig.~\ref{fig3}(b)).
The spectrum above threshold is displayed in Fig.~\ref{fig3}(g): it shows narrow linewidth laser emission from the top of the $sp$ band at about 1454.5~meV (measured linewidth $\sim 60~\mu\mathrm{eV}$, which is the spectrometer resolution), and from the isolated $s$ modes of the B pillars under the excitation spot, each of them emitting at slightly different energies.
The real-space emission at the energy of the lasing $sp$ mode displayed in Fig.~\ref{fig3}(h) reveals the antibonding nature of the modes, with nodes at the junctions between the $s$ and $p$ lobes.
Remarkably, this laser mode is constrained to a single line and presents coherent emission all over its length see Supplemental Material).

The observed blueshift between the top of the $sp$ bands at low power (Fig.~\ref{fig3}(e), dashed line) and the energy of the lasing mode above threshold (Fig.~\ref{fig3}(g)) arises from the loss of strong coupling at high powers.
Indeed, at the low temperatures of the experiments, the exciton resonance of the quantum well at 1460.4~meV and the photonic bands are in strong light-matter coupling, with a characteristic Rabi splitting of about 3.5~meV, as measured in a similar sample~\cite{blessan_directional_2025}.
At low power, the strong coupling shifts down the energy of the upper $sp$ modes by about 0.3~meV with respect to the bare photonic modes.
At high excitation power, at which $sp$ lasing takes place, phase-space filling effects in our cavity with a single quantum well induce the loss of strong coupling and lasing takes place at the energy of the bare photonic modes~\cite{Butte2002a}.
The measured shift in energy of the upper $sp$ mmode in the pumped line is displayed in Fig.~\ref{fig3}(c) and shows a saturation for excitation powers above 10~mW, which is compatible with the loss of strong coupling.

The blueshift of the pumped line can be pictured as a local optical potential with important consequences for the polarization of the lasing mode. 
The effective lateral confinement induced by the local potential results in a splitting of the modes linearly polarized along and perpendicular to the pumped line, see Fig.~\ref{fig3}(c).
This polarization splitting appears naturally in any elongated structure due to the different penetration lengths of the electric field oriented parallel and perpendicular to the optical potential boundary~\cite{gayral_optical_1998,Wertz2010,Tanese2012,rozas_effects_2021, widmann_artificial_2026}.
As the transverse spread of the line modes depends on their linear polarization, modes with different polarizations may couple differently to the excitation spot and have different gains.
In this case, lasing takes place in the vertically polarized mode (oriented perpendicular to the excited line, Fig.~\ref{fig3}(b) - blue dots),
which is not the mode with the highest energy (the horizontally polarized is), but it is the first to lase among the $sp$ modes.
Figure~\ref{fig3}(d) displays the degree of vertical polarization of the mode defined as:
\begin{equation}
\mathrm{DOVP} = \frac{I_{\mathrm{V}} - I_{\mathrm{H}}}{I_{\mathrm{V}} + I_{\mathrm{H}}}.
\label{eq:DOLP definition}
\end{equation}
In this equation, \( I_{\mathrm{H}} \) and \( I_{\mathrm{V}} \) are the emitted intensities measured in the horizontal and vertical linear polarization basis, respectively, at the energy of the lasing mode.
Above threshold, the DOVP of the lasing mode remains constant and very close to 1.

%%%%%%%%%%%%%%%%%%%%%%%%%%%%%%%%%%%%%%%%%
\begin{figure*}[t!]  
\centering
\includegraphics[width=\textwidth]{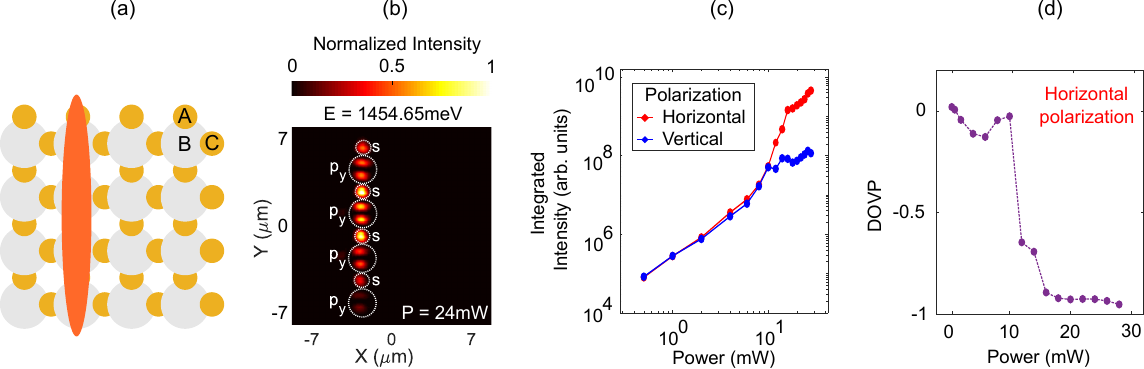}
\caption{(a) Sketch of the excitation configuration of a vertical line of the lattice. (b) Real-space emission at the energy of the $sp$ lasing mode above threshold. (c) Integrated intensity of the upper $sp$ mode along the line in which the lattice is excited. The intensity is resolved in linear polarization. (d) Degree of vertical polarization computed from (c).
%Characterization of the laser emission similar to Fig.~\ref{fig3} for an excitation spot in the vertical direction. 
\label{fig5}}
\end{figure*}
%%%%%%%%%%%%%%%%%%%%%%%%%%%%%%%%%%%%%%

Similar results are observed for an excitation spot aligned along any other of the horizontal and vertical lines of the orbital Lieb lattice. 
An example of lasing under excitation along a vertical line is shown in Fig.~\ref{fig5}(b). 
In this case, the lasing emission is polarized in the horizontal direction (perpendicular to the excited line, see Figs.~\ref{fig5}(c)-(d)), which is consistent with the observations of Fig.~\ref{fig3}. 
The same behavior is found experimentally in other $sp$ Lieb lattices fabricated on the same wafer (see Supplemental Material).

\section{Phase locking of crossing laser modes}
\begin{figure*}[t!]
\centering\includegraphics[width=0.8\textwidth]{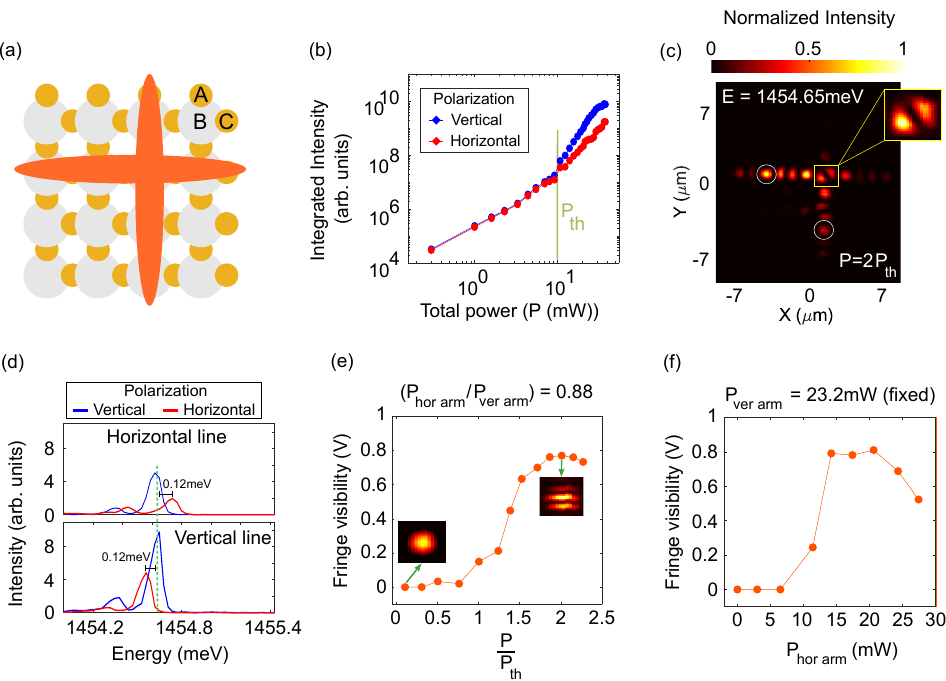}
\caption{(a) Scheme of the crossed excitation spot along a vertical and a horizontal line. (b) Integrated intensity of the top of the $sp$ band integrated over the sites along the lines in which the lattice is excited. (c) Measured laser emission at the energy of the $sp$ lasing mode above threshold ($P=2P_{th}$). The inset shows a zoomed view of the micropillar at which the two lines cross. (d) Measured spectrum resolved in linear polarization at $P=1.2P_{th}$ for photon energies at the top of $sp$ band at two different spatial points corresponding to the circles in the horizontal and vertical arms in (c). (e) Visibility of the fringes arising from the interference between the two circled sites in (c) when they are overlapped on a CCD camera. Insets show the overlapped images at low and high power. (f) Same as (e) when the power of the vertical excitation spot is kept at 23.2~mW and the power of the horizontal excitation spot is varied.} \label{fig6}
\end{figure*}

To further demonstrate the versatility of the orbital Lieb lattice to engineer intricate laser modes, we study the possibility of implementing a laser mode in the form of two crossing lines. 
To do so, we engineer two identical elongated excitation spots that cross at a large diameter B micropillar, see scheme in Fig.~\ref{fig6}(a). To avoid interference effects at the crossing site, the two excitation spots have perpendicular linear polarizations.

Figure~\ref{fig6}(b) displays the emission measured at the energy of the top of the $sp$ band integrated spatially across the two excited lines when the total excitation power is increased. 
For reasons that will be explained below, the intensity of the horizontal pump spot is 0.88 times that of the vertical spot. 
We observe a lasing threshold $P_{th}$ of 10~mW, with dominating emission in vertical polarization.
Figure~\ref{fig6}(c) displays the measured spatial pattern of the laser emission at $2P_{th}$: it shows lasing in the form of a cross under the excitation spots.

The two crossing lasing lines emit at the same photon energy, are phase-locked and share the same polarization (vertical).

Let us first characterize the observed energy and phase locking of the two lines.
This locking is clearly evidenced in the inset of Fig.~\ref{fig6}(c), which displays a zoom of the B pillar at which the two lines cross. 
The intensity pattern with diagonal lobes corresponds to the specific linear combination of $|p_x \rangle + |p_y \rangle$ modes in that pillar.
A random relative phase between the two lasing modes or a difference in energy would result in the averaging of many different relative phases between the $p_x$ and $p_y$ lobes in our time integrated experiments, and the pattern at the crossing micropillar would rather show a ring-like shape. 
Such ring-like pattern is actually observed in the low power regime of spontaneous emission (see Supplemental Material).

%%%%%%%%%%%%%%%%%%%%%%%%%%%%%%%%%%%%%%%%%
%%%%%%%%%%%%%%%%%%%%%%%%%%%%%%%%%%%%%%%%%
%%%%%%%%%%%%%%%%%%%%%%%%%%%%%%%%%%%%%%%%%%
%%%%%%%%%%%%%%%%%%%%%%%%%%%%%%%%%%%%%%%%%%

As an additional proof of the energy and phase locking between the lasers, we study the interference pattern when the emission from the two circled micropillars in Fig.~\ref{fig6}(c) are overlapped in a CCD camera using a set of beamsplitters and mirrors (see Supplemental Material for the details of the interferometric set-up).
Figure~\ref{fig6}(e) shows the measured visibility of the interference fringes of the overlapping spots as a function of the total pump power, which is defined as:
\begin{equation}
    \text{visibility} = \frac{I_{\max} - I_{\min}}{I_{\max} + I_{\min}},
\end{equation}
where $I_{\max}$ and $I_{\min}$ are the maxima and minima intensities around the central interference peak.
At low power, in the spontaneous emission regime (see the inset at $P/P_{th}=0.1$), no fringes are observed in the two overlapping spots in the CCD, resulting in visibility close to zero.
At high power, the observation of deep stable fringes with high visibility (see the inset at $P/P_{th}=2$) in our continuous wave experiment with integration times of the order of several seconds confirms the robust energy and phase locking of the two lasing energy lines.

We now address the conditions for polarization locking in the crossed lines laser.
Figure~\ref{fig6}(d) displays the measured spectrum resolved in linear polarization at $P=1.2P_{th}$ for photon energies at the top of $sp$ band. 
The measurements are done at two different spatial points corresponding to the circles in the horizontal and vertical arms in Fig.~\ref{fig6}(c).
For the horizontal lasing line, the top of the $sp$ band is split in modes with perpendicular linear polarizations. As discussed above for a single line excitation, the origin of this splitting is related to the blueshift of the lasing line at high excitation powers.
The top panel of Fig.~\ref{fig6}(d) shows explicitly this splitting.
The majority of the emission is in the lasing mode with vertical polarization (blue line), which emits at lower energy than the horizontal polarization (red line). 
Note that at higher pump powers, the vertical polarized intensity largely overwhelms the horizontally polarized emission, the power in the spectra displayed in Fig.~\ref{fig6}(d) was chosen such that the polarization splitting could be easily visible.
If we now focus on the spectra emitted by the vertical line, we observe the same hierarchy of splittings but now rotated by $90^{\circ}$:
the majority of the emission is also in the vertical polarization (blue line), but now this mode is at higher energy than the horizontally polarized emission (red line).
The splitting is 0.12~meV in both lines.

We therefore see that the blueshifts induced by the loss of strong coupling under the pump spots result in the splitting of the polarization in the vertical and horizontal lines and are such that the vertically polarized modes have the same photon energy in both lines.
To achieve this energy degeneracy, each arm requires a different blueshift, which in turn requires different pump powers. %This is the reason why the excitation spots are arranged with different powers: $P_{\text{hor arm}}=0.88P_{\text{ver arm}}$.
%Note that the energy of the lasing horizontal line in Fig.~\ref{fig6}(d) is higher than the one of the vertical line. 
In particular, the emission in the horizontal line shows a larger blueshift, indicating that more carriers are injected in this line than in the vertical one (the blueshift is proportional to the local carrier density). 
However, the measured excitation power before the sample was stronger in the vertical spot than in the horizontal spot ($P_{\text{hor arm}}=0.88P_{\text{ver arm}}$).
This discrepancy may be caused by deviations from perfect alignment of the two pump spots, which resulted in a worse line overlap for the vertical spot.
The Supplemental Material shows a situation in which the cross laser lases in the horizontal polarization.
To reach this situation, the imbalance of intensities of the pumps is set such that energy degeneracy between the two lines is reached for the horizontal polarization at threshold.

To get deeper insights into the energy, phase and polarization locking between the two laser lines, we have tested the resilience of the fringe visibility to asymmetries in the excitation power of the two lines. 
To do so, we increase the power of horizontal excitation spot while the power of the vertical spot is kept at 23.2~mW. 
The measured fringe visibility is plotted in Fig.~\ref{fig6}(f). 
Only when the two arms have similar powers the fringe visibility is high, witnessing locking between the two lasers.

\section{Numerical model of lasing}
To investigate the mechanisms of phase locking between the crossing line lasers observed experimentally, we numerically model the lasing emission of a Lieb-$sp$ lattice of $4 \times 4$ unit cells.
As a first step, we neglect the polarization degree of freedom and consider scalar fields.
While this simplification is significant, it provides relevant insights on the conditions for phase locking.
More insights on the the polarization degree of freedom are discussed at the end of this section.

Let us focus on the $sp$ bands and ignore lasing in the $s$ modes of the B pillars, which is systematically observed along lasing in the line modes of the $sp$ bands.
The time evolution of the scalar photon field $\{|\psi_{l,m}^{J,\mu}\rangle\}$ at unit cell $l,m$, site J=A,B,C and orbital $\mu=s,p_x,p_y$ is governed by the following differential equation~\cite{Carusotto2013}:
%%%%%%%%%%%%%%%%%%%%%%%%%
\begin{widetext}
\begin{equation}
i\hbar\frac{d}{dt}|\psi_{l,m}^{\text{J},\mu}\rangle = \left( H_{\text{sp}}+H_{p_x p_y} - i\frac{\Gamma_p}{2} + g_R n_{l,m}^\text{J}  +i \frac{\Gamma_d}{2}n_{l,m}^J\right) |\psi_{l,m}^{\text{J},\mu}\rangle + \frac{\sqrt{n_{l,m}^{\text{J}}}}{2} \Gamma_d e^{i\phi_{l,m}^{\text{J},\mu}(t)}.
\label{eq:psi_evolution}
\end{equation}
\end{widetext}
%%%%%%%%%%%%%%%%%%%%%%%%%
\noindent $H_{\text{sp}}$ is the tight-binding Hamiltonian Eq.~(\ref{eq:H_sp}). $H_{p_xp_y}$ is a term that couples the $p_x$ and $p_y$ orbitals of each individual micropillar B with strength $-t_{p_xp_y}$. 
As we will see below, this term is crucial to reproduce the phase locking in the cross laser. 
\(\Gamma_p = 62.5\,\mu\mathrm{eV}\) is the decay rate via escape of photons from the cavity. This value is estimated from an independent measurement of the photon lifetime on a one-dimensional lattice in the same wafer~\cite{blessan_directional_2025}. 
The next three terms in Eq.~(\ref{eq:psi_evolution}) involve the coupling of the photon field to an exciton reservoir fed by the pump laser. 
\( g_R\) is the strength of the nonlinear interaction between the photon field and the reservoir, which results in a blueshift of the onsite photon energies. As mentioned above, this blueshift is a consequence of the loss of strong coupling due to phase space filling. \(\Gamma_d\) is the gain from the reservoir into the photon field.
The last term is a source term that accounts for the spontaneous emission from the gain medium (the exciton reservoir), in which \(\phi_{l,m}^{\text{J},\mu}(t)\) is a random phase uniformly distributed in $[0, 2\pi)$ and serves as a seed for lasing. 

The photon field is coupled to the exciton reservoir $n_{l,m}^{\text{J}}$ (active gain medium) at each site $\{l,m\}$ pumped by the external laser. The $p_x$ and $p_y$ modes of the B sites share a common reservoir whose dynamics is given by:
\begin{equation}
\hbar\frac{dn_{l,m}^\text{B}}{dt} = 2P_{l,m}^\text{B} - \Gamma_e{n_{l,m}^\text{B}} - \Gamma_d{n_{l,m}^\text{B}}\left( |\psi_{l,m}^{\text{B},p_x}|^2 +1+ |\psi_{l,m}^{\text{B},p_y}|^2 + 1 \right).
\label{eq:n_shared}
\end{equation}

\noindent The $\hbar$ prefactor on the left hand side has been added for consistency of units. The $s$ modes in the A and C micropillars have individual reservoirs:
\begin{equation}
\hbar\frac{dn_{l,m}^\text{J}}{dt} = P_{l,m}^\text{J} - \Gamma_e n_{l,m}^\text{J} - \Gamma_d n_{l,m}^\text{J}\left(|\psi_{l,m}^{\text{J},s}|^2 + 1 \right).
\label{eq:n_unshared}
\end{equation}

\noindent The pump profile is given by $P_{l,m}^\text{J}$. The factor of 2 in Eq.~(\ref{eq:n_shared}) is a geometric factor: It accounts for the larger size of the B micropillars whose reservoir is about twice as large as for the A and C pillars.
\(\Gamma_e\) is the decay rate of the reservoir.
As mentioned above, $\Gamma_d$ is the emission rate into the photon field. 
The values of the parameters used in the simulations are \(\Gamma_e = 0.417\,\mu\mathrm{eV}\) and  $\Gamma_d= 1.52\,\mu\mathrm{eV}$, which fall in the typical range of parameters used in micropillar lattices under non-resonant excitation~\cite{Baboux2018, Fontaine2022}.
The pump profile $P_{l,m}^\text{J}$ is implemented as two discretized elongated Gaussian spots applied along the horizontal and vertical lines of the lattice corresponding to the experimental conditions of Fig.~\ref{fig6}. 
At each site, the pump intensity is assigned based on its position along the Gaussian envelope. 
%In the experiment, the two crossing excitation spots have perpendicular polarizations.
The site at which the horizontal and vertical spots intersect receives contributions from both spots.
The time evolution of the photon field and the reservoir populations are computed using the ode45 solver, which implements an explicit Runge-Kutta (4,5) method based on the Dormand–Prince pair.
This solver uses adaptive step sizing to efficiently and accurately integrate coupled nonlinear differential equations.
We compute the photon fields and reservoir densities in the steady state at long times after switching on the laser excitation $P_{l,m}^\text{J}$.

In our experiment, lasing consistently occurs in line modes at the top of the $sp$ antibonding band, regardless of whether we use a single elongated pump spot or two crossed spots. 
This preference for the antibonding band edge, also observed in previous microcavity lattice studies~\cite{Lai2007b, Tanese2013, Jacqmin2014}, is due to its longer photon lifetime. 
The spatial profile of these antibonding modes has a node at the junctions between adjacent pillars. 
These junctions have the smallest etching features of the structure and the highest density of non-radiative defects. 
In contrast, the wavefunctions of the lower-energy bonding modes exhibit high amplitude at these very junctions, subjecting them to stronger dissipation and making them less favorable for lasing. 
To model this feature, we have introduced a dissipative coupling term~\cite{Ohadi2015} in $H_{\text{sp}}$ of Eq.~(\ref{eq:H_sp}), by replacing $-t_{\text{sp}}$ with $-t_{\text{sp}}-i\gamma_{\text{sp}}$. 
This imaginary hopping term results in complex eigenvalues of  $H_{\text{sp}}$ whose imaginary part can be associated to photon losses and is smaller for the upper part of the band of eigenvalues than for the lower part.
Using a value of $\gamma_{\text{sp}}$  as small as $5\,\mu\text{eV}$, our simulations systematically produce lasing at the top modes of the $sp$ band. 

However, dissipative coupling in combination with the cross-shaped pump profile is not enough to reproduce the observed phase locking between horizontal and vertical lines. 
Simulations in these conditions result in simultaneous lasing in the horizontal and vertical lines at the energy of the top of the band, but their relative phase is random at each simulation realization.
To reproduce the phase locking observed in the experiment, in Eq.~(\ref{eq:psi_evolution}) we incorporate a coupling term between the $p_x$ and $p_y$ orbitals of each B site: $H_{p_xp_y}=\sum_{\text{B sites}}{ -t_{p_xp_y}|\psi_{l,m}^{\text{B},p_x}\rangle\langle\psi_{l,m}^{\text{B},p_y}| + h.c.}$
A possible origin of this term is a residual ellipticity in the shape of the B pillars, most probably introduced during the fabrication at the electron beam lithography or reactive ion etching processes.
This ellipticity breaks the degeneracy between the ${p_x}$ and ${p_y}$ orbitals. 
Through numerical simulations across 100 independent stochastic realizations, we find that for the parameters used in our simulations, a minimum coupling strength $t_{p_xp_y} = {37.5\,\mu\text{eV}}$ (nearly half of $\Gamma_p$) is required to induce phase locking of the two line modes with the observed relative phase.

Based on studies of the splitting between $p_x$ and $p_y$ modes in micropillars with large ellipticities~\cite{sebald_optical_2011, galimov_towards_2023}, we estimate that the ratio of the short to the long axis of our micropillars required to reach the $p_x-p_y$ coupling used in the simulations should be of the order of 0.994.
This ellipticity is too close to 1 (perfectly circular micropillar) to be visible in a scanning electron microscope image, in particular given the spatial overlap the adjacent micropillars.
It is also too small to be revealed in the photoluminescence of the real space and far field emission at low excitation power measured in Fig.~\ref{fig2}(b)-(d): the coupling between $s$ and $p$ modes is more than 10 times larger than $t_{p_xp_y}$.
Only in the crossing lines lasing configuration it plays an important role in the phase locking of the two lasers.

\begin{figure*}[t!]
\centering\includegraphics[width=0.85\textwidth]{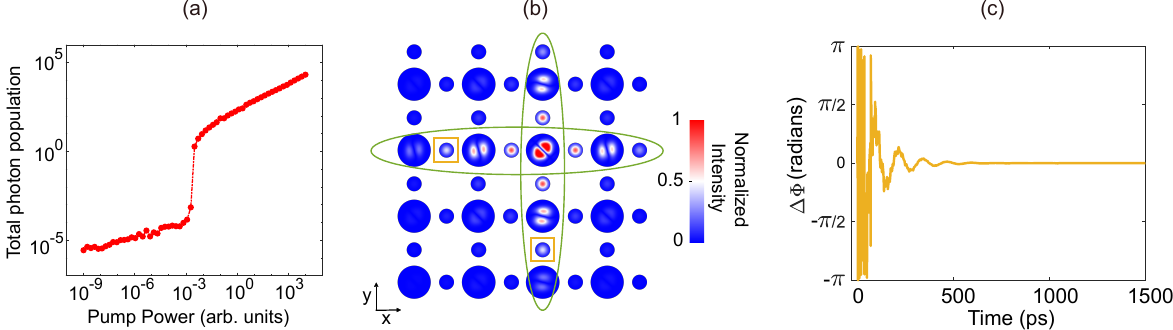}
\caption{Numerical simulations in the cross pump configuration. (a) Total photon population as a function of pump power, showing the onset of lasing.
(b) Real-space intensity profile of the lasing mode above the lasing threshold, illustrating line lasing within the lattice.
(c) Relative phase difference over time between two selected sites, one in the horizontal arm and one in the vertical arm highlighted by rectangles in (b). The fixed value of the relative phase at long times demonstrates phase locking between emission from the two lines.\label{fig7}}
\end{figure*}

Figure~\ref{fig7} provides an illustration of the emergence of lasing and phase-locking in the simulated scalar Lieb-$sp$ lattice with the excitation conditions of the experiments reported in Fig.~\ref{fig6}. 
Panel (a) displays the total photon population numerically computed as a function of the applied pump power in log-log scale. For each pump power, the total photon population is evaluated at long enough times for the system to reach the steady state. The characteristic threshold behavior is evident, with a sharp increase in photon population once the pump exceeds a critical value. This point marks the onset of macroscopic occupation, signaling the transition to the lasing regime.
Panel (b) displays the steady state spatial distribution of the lasing mode in real space, taken at a pump value well above threshold. 
%This spatial distribution is obtained by projecting the final condensate wavefunction at the steady state onto the eigenmodes of the bare Hamiltonian (no \(\gamma_{sp}\), \(t_{p_xp_y}\) and \( g_R\)). This projection reveals which eigenmodes dominate the condensate. From the lasing eigenmode, we extract both amplitude $A_n = |\psi_n|$ (the strength of the mode on each orbital) and the phase $\phi_n = \arg(\psi_n)$ associated with each orbital. These quantities are stored for all 64 orbitals and used for visual reconstruction. 
To visualize the computed lasing pattern in real space, we assign spatial profiles to each orbital type: \(p_x\) and \(p_y\) orbitals are represented by antisymmetric Gaussian lobes; \(s\) orbitals are represented by symmetric Gaussian spots in the small pillars.

Figure~\ref{fig7}(b) reproduces qualitatively the phase-locked crossed lasers observed in the experiment.
The relative phase of the $p_x$ and $p_y$ modes in the simulation matches the observed one, with an orientation of the $p$ lobes at the crossing site corresponding to the linear superposition $|p_x\rangle + |p_y\rangle$.
Note that if we change the sign of $t_{p_xp_y}$ in the simulations, the locking phase is turned by $\pi$ (i.e., $|p_x\rangle - |p_y\rangle$).

To get further insights into the phase locking dynamics we analyze the simulated time evolution of the relative phase of the light field at two sites after switching on the pump laser: one located along the horizontal arm and the other along the vertical arm, both highlighted by a rectangle in Fig.~\ref{fig7}(b).
At short times (up to $\sim 100\,\text{ps}$), spontaneous emission from the reservoir into the lattice modes results in a random phase evolution in both sites, indicating the absence of long-range phase coherence.
When occupation of the lasing mode starts to take over ($\sim 100-300,\text{ps}$), the relative phase displays coherent oscillations which end up in phase locking at longer times.

Let us now discuss the polarization degree of freedom.
A complete simulation of the polarization dynamics would require an elaborate model with a significant number of fitting parameters.
In particular, simulating the blueshifts and polarization splitting as a function of the pump power in each line requires models beyond what is available in the literature.
Nevertheless, we can test our description of the polarization locking of the two lines based on the energy degeneracy of same polarization modes in the two crossing lines.
To do so, in the Supplemental Material, we detail a simulation of the cross line configuration in the lasing regime including the experimentally measured polarization splitting of each line observed in the experiment and varying the blueshifts of each respective line.
When the blueshifts are such that the same polarization is aligned in energy in both lines, we do see lasing with phase and polarization locking.
When the blueshifts are such that they do not align equal polarization modes, each line lases in a different polarization.
These simulations confirm the key role of the energy alignment of different polarizations through the appropriate blueshifts.

\section{Conclusions}
The combination of different orbitals in lattices of photonic resonators is a useful resource to engineer unconventional photonic band structures.
Here we have used this feature to implement a two-dimensional Lieb-$sp$ lattice with line modes that can lase along any of the main lines of its square geometry.
When the gain is designed in the form of two crossing lines, we have shown that phase and polarization locking between the two line modes is possible.
The two features (independent lasing of parallel lines and locking of crossing lines) anticipate great flexibility in the design of densely integrated lasers with specialty real-space patterns, particularly when considering electrical injection with line or site resolution.
First observations of lasing under electrical injection in lattices of coupled micropillar anticipate exciting prospects in this directions~\cite{Suchomel2018, suchomel_spatio-temporal_2020}.
The use of novel cavity polariton materials such as those based on GaN and ZnO semiconductors, perovskites, transition metal dichalcogenides and different types of organic materials which operate in the strong coupling regime at room temperature~\cite{ghosh_microcavity_2022}, are also promising to design this kind of lasers in operation at ambient conditions.

\textit{Acknowledgements.}
This project funded within the QuantERA II Programme that has received funding from the EU H2020 research and innovation. It has also been supported by the European Union’s Horizon 2020 research and innovation programme through the ERC projects EmergenTopo (grant number no. 865151) and ARQADIA (grant agreement no. 949730); the Marie Sklodowska-Curie grant agreement no.101108433; and under Horizon Europe research and innovation programme  through the ERC project ANAPOLIS (grant agreement no. 101054448).
It was also funded by the French government through the Programme Investissement d'Avenir (I-SITE ULNE /ANR-16-IDEX-0004 ULNE) managed by the Agence Nationale de la Recherche, the Labex CEMPI (ANR-11-LABX-0007), the CDP C2EMPI project (R-CDP-24-004-C2EMPI), as well as the French State under the France-2030 programme, the University of Lille, the Initiative of Excellence of the University of Lille, the European Metropolis of Lille, and the region Hauts-de-France via CPER Wavetech. It was partly supported by the Paris Ile de France R\'egion in the framework of DIM SIRTEQ and DIM QuanTIP and by the RENATECH network and the General Council of Essonne.

\textit{Data availability}
Data underlying the results presented in this paper are available in Ref.~\cite{data_availability}.

\textit{Supplemental Material}
Supplemental Material accompanies this work as an ancillary file to the arXiv submission. It can be found at \url{https://arxiv.org/abs/2512.18719}

\textit{Disclosure}
The authors declare no conflicts of interest.

%%%%%%%%%%%%%%%%%%%%%%% References %%%%%%%%%%%%%%%%%%%%%%%%%
%\subsection*{References}
%\bibliography{biblio.bib}

\begin{thebibliography}{48}
\bibitem [{dat()}]{data_availability}%
  \BibitemOpen
  \href@noop {} {\bibinfo {title} {The data sets correspoding to the figures of
  this article are available at {DOI}: 10.57745/qxi7gr}}\BibitemShut {NoStop}%
\end{thebibliography}
%\input{main_05_arxiv.bbl}
%

\end{document}